\documentclass[fleqn,usenatbib]{mnras}

\usepackage{newtxtext,newtxmath}

\usepackage[T1]{fontenc}
\usepackage{ae,aecompl}

\usepackage{graphicx}	
\usepackage{amsmath}	
\usepackage{amssymb}	
\usepackage{soul}

\title[Not Gone with the Wind]{Not Gone with the Wind: Planet Occurrence is Independent of Stellar Galactocentric Velocity}

\author[McTier \& Kipping]{
Moiya A.S. McTier$^{1}$\thanks{E-mail: mmctier@astro.columbia.edu} and
David M. Kipping$^{1}$
\\
$^{1}$Department of Astronomy, Columbia University, 550 W 120th St., New York, NY 10027\\
}

\date{Accepted XXX. Received YYY; in original form ZZZ}

\pubyear{2019}

\begin{document}
\label{firstpage}
\pagerange{\pageref{firstpage}--\pageref{lastpage}}
\maketitle

\begin{abstract}
We demonstrate that planet occurrence does not depend on stellar galactocentric velocity in the Solar neighborhood. Using \textit{Gaia} DR2 astrometry and radial velocity data, we calculate 3D galactocentric velocities for 197,090 \textit{Kepler} field stars, 1647 of which are confirmed planet hosts. When we compare the galactocentric velocities of planet hosts to those of the entire field star sample, we observe a statistically significant (KS p-value = $10^{-70}$) distinction, with planet hosts being apparently slower than field stars by $\sim$40 km/s. We explore some potential explanations for this difference and conclude that it is not a consequence of the planet-metallicity relation or distinctions in the samples' thin/thick disk membership, but rather an artefact of \textit{Kepler}'s selection function. Non \textit{Kepler}-host stars that have nearly identical distances, temperatures, surface gravities, and \textit{Kepler} magnitudes to the confirmed planet hosts also have nearly identical velocity distributions. Using one of these identical non-host samples, we consider that the probability of a star with velocity $v_{\mathrm{tot}}$ hosting a planet can be described by an exponential function proportional to $e^{(-v_{\mathrm{tot}}/v_0)}$. Using a Markov Chain Monte Carlo sampler, we determine that $v_0 > $976 km/s to 99\% confidence, which implies that planets in the Solar neighborhood are just as likely to form around high-velocity stars as they are around low-velocity stars. Our work highlights the subtle ways in which selection biases can create strong correlations without physical underpinnings.
\end{abstract}




\section{Introduction}
Since the discovery of the first exoplanet more than two decades ago, astronomers have found nearly 4000 planets outside of our solar system (\href{https://exoplanetarchive.ipac.caltech.edu/}{NASA Exoplanet Archive}; \citet{akeson:2013}). As far as we can tell, these planetary systems are incredibly diverse, ranging in multiplicity, orbital parameters, planet size, etc. But one thing we've found to be common amongst these systems is that properties of the host star influence characteristics of orbiting planets. For example, we know that massive planets are more common around metal-rich stars and that smaller planets are more common around low-mass stars \citep{mulders:2018}. Another classic example is that metal-rich stars appear to more frequently host hot Jupiter companions \citep{wang:2015}.

Chemical abundance surveys and exquisite astrometric datasets like \textit{Gaia} have shown that stellar populations aren't distributed homogeneously throughout the Milky Way. Instead, astronomers have observed gradients in motion \citep{trick:2019}, metallicity \citep{toyouchi:2018,lemasle:2018}, and stellar number density \citep{valenti:2016,mcgaugh:2016} that can all be attributed to galactic motion and structure.

Given that planet characteristics are closely tied to stellar properties and that stellar properties trend with galactic behavior, it's not unreasonable to search for a connection between exoplanet traits and galactic evolution. \citet{gonzalez:2001} attempted to do this in terms of galactic chemical evolution by linking the mass of a planet to the metallicity of the gas cloud from which its host star formed. They determined that only regions of the galaxy with stellar metallicity exceeding half of the Sun's could produce habitable planets. This criterion for habitability was later used to constrain a habitable zone for M31 \citep{carigi:2013}. In addition to sufficient metallicity, \citet{lineweaver:2004} claimed that habitable planets must be in a region of the galaxy with low levels of harmful radiation from supernovae. \citet{sundin:2006} added that the speed of a host star's orbit is an important factor in galactic habitability for spiral galaxies and studied how the presence of a bar could complicate stars' orbits to the point where they are no longer suitable planet hosts. Stellar dynamics has also been the focus of more recent work with studies examining the survival rates of planets in dense stellar cluster environments \citep{dotti:2018, fujii:2019} due to the fact that close stellar encounters can strip away planets or destabilize their orbits \citep{elteren:2019,daohai:2019}. Much of this work was theoretical or completed before the advent of the \textit{Kepler} planet catalogue. 

Our goal in this work was to take a step back from the search for habitable planets in a galactic context to determine if there are galactic conditions under which planets are more likely to form. After all, you can't find a habitable planet where there are no planets to begin with. To that end, we took advantage of \textit{Gaia}'s extensive astrometric data alongside the \textit{Kepler} exoplanet database to observationally explore the relationship between planet occurrence rate and stellar motion throughout the Milky Way. 

The structure of this paper is as follows. In \S~\ref{sec:data}, we describe the datasets we use and our sample selection process. We describe our observation and explore possible explanations in \S~\ref{sec:obs}. Finally, we summarize our work and discuss opportunities for future research in \S~\ref{sec:summary}.

\section{Data \& Sample Selection}\label{sec:data}
To study the relationship between stellar motion and planet occurrence rate, we used a cross matched catalog of astrometric data from \textit{Gaia} DR2 \citep{gaia:2018} and exoplanet data from \textit{Kepler} DR25 \citep{thompson:2018}. We also used actions calculated by \citet{sanders:2018} to better understand stellar motion in a wider galactic context. These data are described further in the following subsections.

\subsection{\textit{Gaia} Astrometry}\label{sec:gaia}
\textit{Gaia} is an all-sky survey whose primary science goal is to measure the spatial and velocity distributions of stars in the Milky Way and determine their astrophysical properties. 

The second \textit{Gaia} data release \citep{gaia:2018} includes five astrometric parameters -- positions, parallax, and proper motions -- for more than 1.3 billion stars. Unlike the first \textit{Gaia} data release, the parallaxes derived in this latest dataset are based only on \textit{Gaia} observations and have not had to rely on information from the Tycho-2 Catalogue. DR2 also includes radial velocity (RV) data for more than 7.2 million stars with mean G magnitudes between 4 and 13 and effective temperatures (T$_{\mathrm{eff}}$) between 3550 K and 6900 K.

We used these astrometric parameters to calculate galactocentric velocities for our sample. We describe the calculation in \S~\ref{sec:obs}.

For \textit{Gaia} DR2, parallax uncertainties at the faint end of the Gaia magnitude range (G = 20 mag) are $\sim$0.7 milliarcseconds (mas). Proper motion uncertainties at this faint end are $\sim$1.2 mas/yr. Radial velocities at this faint end have errors of 1.2 km/s for stars with temperatures of 4750 K while stars with temperatures of 6500 K can expect RV uncertainties of $\sim$2.5 km/s. 

These large uncertainties for faint stars can propagate through our velocity calculations and result in velocity errors >20,000 km/s. 

\subsection{\textit{Kepler} Exoplanet Catalog}\label{sec:kepler}
NASA's \textit{Kepler} Space Telescope was launched in 2009 to study planetary systems \citep{koch:2010}. Using its 0.95m aperture, the spacecraft continuously observed nearly 200,000 main sequence stars in a 105 square-degree field of view situated between the constellations Cygnus and Lyra. \textit{Kepler}'s photometer measured the amount of light received from each star over time, waiting for a dip in brightness that could potentially be caused by a transiting planet.  

The primary \textit{Kepler} mission operated for 4 years until the spacecraft's second reactor wheel failed in 2013. From that data, exoplaneteers have confirmed the existence of 2,342 exoplanets and identified thousands of planet candidate hosts. The spacecraft was recommissioned in 2013 for the \textit{K2} mission, which hunted for transiting planets in 13 different fields along the ecliptic plane. The spacecraft was officially retired in 2018 (\href{https://exoplanetarchive.ipac.caltech.edu/}{NASA Exoplanet Archive}).

In this work, we only considered planet hosts that were observed and discovered by the primary \textit{Kepler} mission. This was done to maximize the number of planets in our sample while avoiding systematic offsets between \textit{Kepler} and \textit{K2}. Inherent in this decision is the assumption that the \textit{Kepler} volume is a representative slice of the Solar neighborhood, but \citet{fernandes:2019} showed that the \textit{Kepler} volume is representative by demonstrating that the planet occurrence rate for giants in the \textit{Kepler} field is similar to that in the entire Solar neighborhood.

\textit{Kepler} observed stars with apparent Kepler magnitudes as bright as 9 and as dim as 16 in an optical bandpass ranging from 420-890 nm. The majority of the mission's target stars were FGK types, with effective temperatures between 3500 and 7500 K \citep{brown:2011}. 

\subsection{Actions}\label{sec:actions}
Following the release of \textit{Gaia} DR2 in April 2018, \citet{sanders:2018} released a catalog of stellar ages for the nearly 3 million \textit{Gaia} stars with radial velocity measurements and reliable spectroscopic data. For the benefit of the community, they also provided complementary galactocentric coordinates and actions. 

Actions are useful when describing orbits because they are known as "integrals of motion." These are quantities that can be described as functions of position and velocity or momentum, and that are conserved throughout the orbit. There are three action coordinates: radial action (J$_{\mathrm{R}}$), vertical action (J$_{\mathrm{Z}}$), and angular momentum (L$_{\mathrm{z}}$). By providing all three action coordinates, one can perfectly describe an entire orbit at any point in time or space.

Traditionally, actions were incredibly difficult to calculate. In fact, they can only be analytically solved for a small subset of convenient galactic potentials. As computers became more advanced, astronomers developed ways to numerically approximate action coordinates.

\citet{sanders:2018} used the St{\"a}ckel Fudge method for approximating actions \citep{binney:2012}, which interpolates on a distribution of phase-space coordinates generated from an axisymmetric potential. Resulting errors on the actions are propagated through from errors on parallax, proper motion, and radial velocity.

We use the vertical actions calculated by \citet{sanders:2018} to determine whether stars in our sample are members of the thick or thin disk (see \S~\ref{sec:disk}).

\subsection{Cross-matching and Quality Cuts}\label{sec:cuts}
In order to compare the motions of our planet host and stellar samples, we used \href{https://gaia-kepler.fun/}{Megan Bedell's catalog} of data cross-matched between \textit{Gaia} DR2 and \textit{Kepler} DR25 with a 1 arcsecond search radius, downloaded on January 9th, 2019. The two catalogs were matched using the RA and Dec coordinates of every star. The final cross-matched catalog contains 197,090 stars, 1,647 of which are confirmed planet hosts. 

We make the following quality cuts on both our planet host and field star samples:

\begin{itemize}
    \item We removed stars with negative parallax values. This left our planet host sample with 1,637 members and our field star sample with 195,545.
    \item We removed giants as well as massive and very-low mass dwarfs from our data by discarding stars with logg < 4 and effective temperatures <4000 or >8000. We did this because \textit{Kepler} was optimized for finding planets around FGK stars and therefore has a lower completeness for giants \citep{sliski:2014}. This left our planet host sample with 1,477 stars and our field star sample with 148,860 stars.
    \item We removed stars with relative errors in the total velocity (see \S~\ref{sec:obs}) >10\% to minimize the error propagated through our analysis. Much of the error in the velocity calculation comes from uncertainties in the parallax, so making this 10\% cut ensures that we're only using stars with small parallax uncertainties and therefore reliable distances consistent with those in \citet{bailer-jones:2018}. This left us with two final samples of \textit{Kepler} FGK dwarfs with reliable galactocentric velocities: a planet host sample with 1,437 members and a field star sample with 140,918 members.
    
\end{itemize}

Similarly, we ensure high data quality for our sample of stars with action values \citep{sanders:2018} by removing giants and stars with errors in vertical action (J$_{\mathrm{Z}}$) <10\%. Because \citet{sanders:2018} only calculated actions for a subset of \textit{Gaia} data, there are 18,102 stars with action coordinates that meet our criteria, 382 of which host transiting planets. 

We refrained from calling our field star sample the non-host sample and removing the transiting planet hosts from it because many of the field stars will surely have some considerable fraction of planets amongst them as planets are common and transits are geometrically rare. 

\begin{figure}
\includegraphics[width=3.2in.]{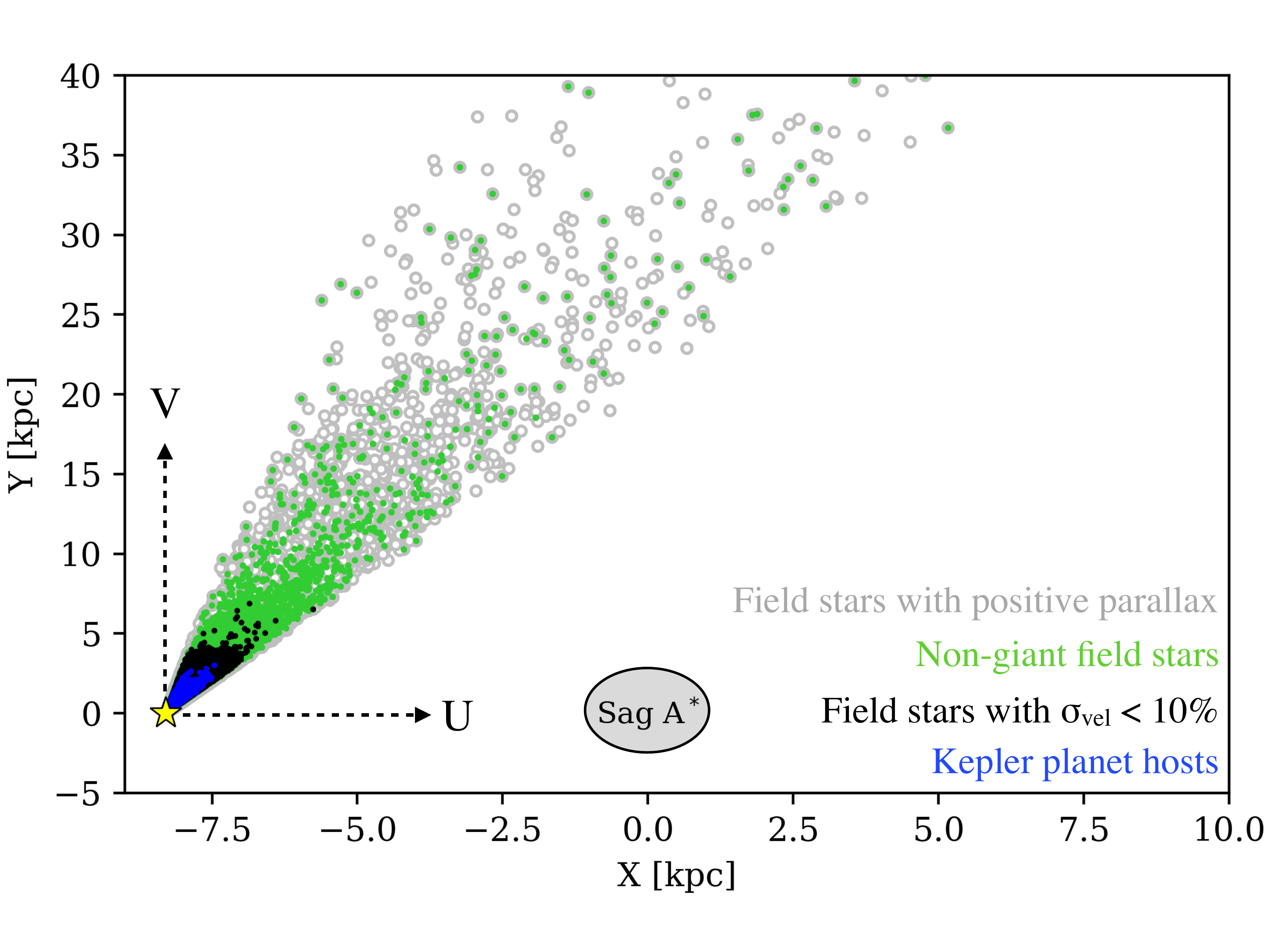}
\caption{Face-on view of galactocentric X-Y positions for the stars in our cross matched catalog. Different colors correspond to different steps in our sample selection process. Grey points show all stars in our cross-matched catalog with physical astrometric values. Green points show that same data with giants removed. Black points show stars with total velocity errors <10\%; these are the members of our \textit{Kepler} field star sample. Blue points show which of those stars are \textit{Kepler} planet hosts.}
\label{fig:data}
\end{figure}


\section{Observation \& Hypothesis Testing}\label{sec:obs}
Having all five astrometric parameters plus RVs from \textit{Gaia} allows us to calculate 3D velocity vectors for all of the stars in our cross matched catalog. We calculated our velocities using $\mathtt{astropy}$ \citep{astropy:2013,astropy:2018}, which performs a series of transformations to translate a star's heliocentric position and motion to galactocentric coordinates and velocities. We repeated the velocity calculation 5000 times, each using parallaxes and proper motions drawn from a normal distribution with their recorded uncertainties. For each velocity calculation, we used median \textit{Gaia} RV measurements instead of drawing from a distribution as there were stars in our crossmatched catalog without reported RV errors. The velocities we used in our analysis are the median velocities of these 5000 trials, and the velocity errors are the standard deviations. As previously mentioned, we then removed from our sample any stars with total velocity errors >10\%.

Our coordinate system is one where U is positive pointing towards the galactic center, V is positive in the direction of the Sun's rotation around the galaxy, and W is positive pointing towards the galactic north pole. We use $\mathtt{astropy}$'s default parameters for the Sun's position and velocity, which assumes a galactocentric radius of 8.3 kpc \citep{gillessen:2009} and a Solar velocity vector of \textbf{v}$_{\odot}$ = [U$_{\odot}$,V$_{\odot}$,W$_{\odot}$] = [11.1,12.24,7.25] km$\:$s$^{-1}$ \citep{schoenrich:2010}. We did not subtract off the local standard of rest (LSR) of 220 km/s.

\begin{figure*}
\includegraphics[width=\textwidth]{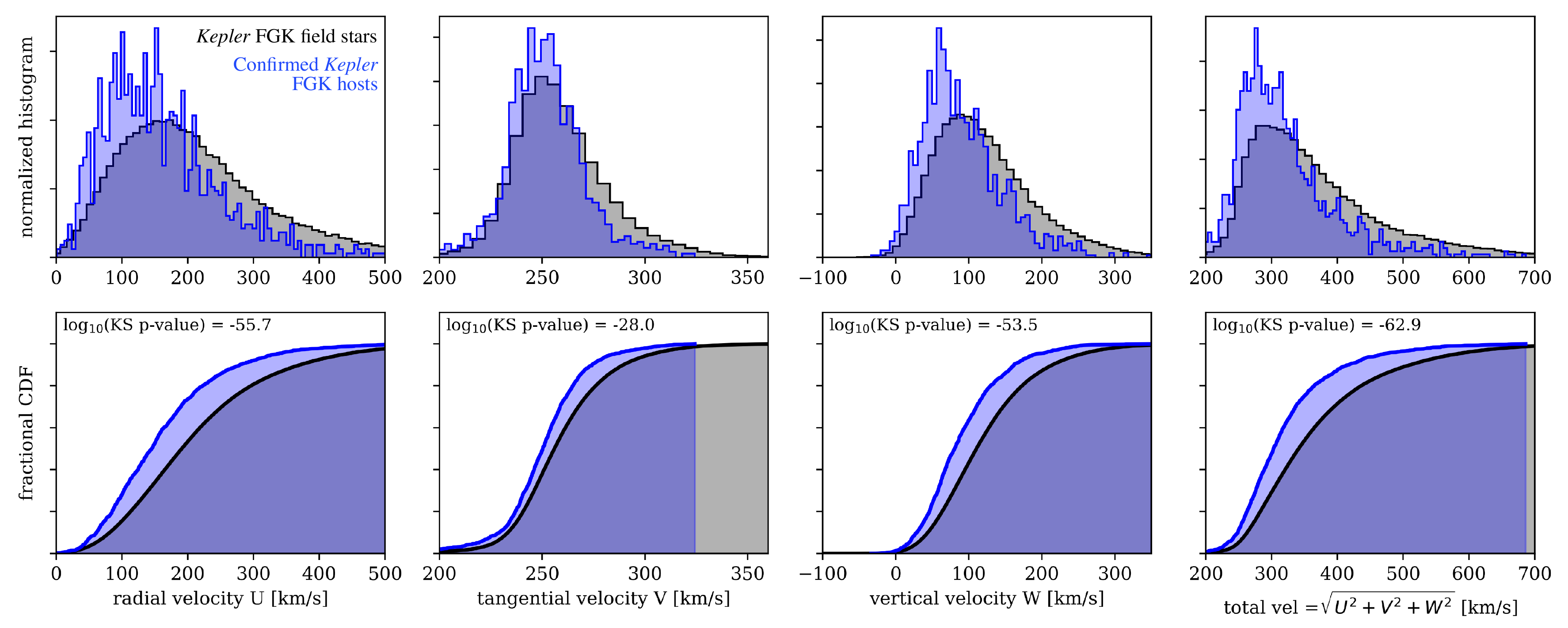}
\caption{\textbf{Top:} Histogram of the galactocentric velocities of \textit{Kepler} planet hosts (blue) versus the ensemble of Kepler field stars (black). \textbf{Bottom:} Same information as in top row shown in cumulative distribution functions (CDFs). Planet hosts move slower than other \textit{Kepler} field stars along every velocity vector to high significance (see KS test p-values inset to each panel).}
\label{fig:mot}
\end{figure*}

Figure~\ref{fig:mot} shows the cumulative distribution functions (CDFs) comparing our two samples in U, V, W, and total velocity ($\sqrt{U^2+V^2+W^2}$). In each panel, you can see that the median velocity for planet hosts is slower than that of our field star sample.

We also statistically compared the two samples by calculating the Kolmogorov-Smirnov (KS) p-value \citep{smirnov:1948} using the $\mathtt{ks\_2samp}$ function from scipy.stats. The 2-sample KS test compares the empirical distribution functions of two samples to test the null hypothesis that they are drawn from the same parent population. A p-value is then  calculated based on the distance between the two empirical distributions. The KS p-value can be interpreted the same as any other so that values below a certain confidence threshold indicate that the null hypothesis should be rejected, i.e. the two samples are not drawn from the same parent population.

\begin{figure}
\includegraphics[width=3in.]{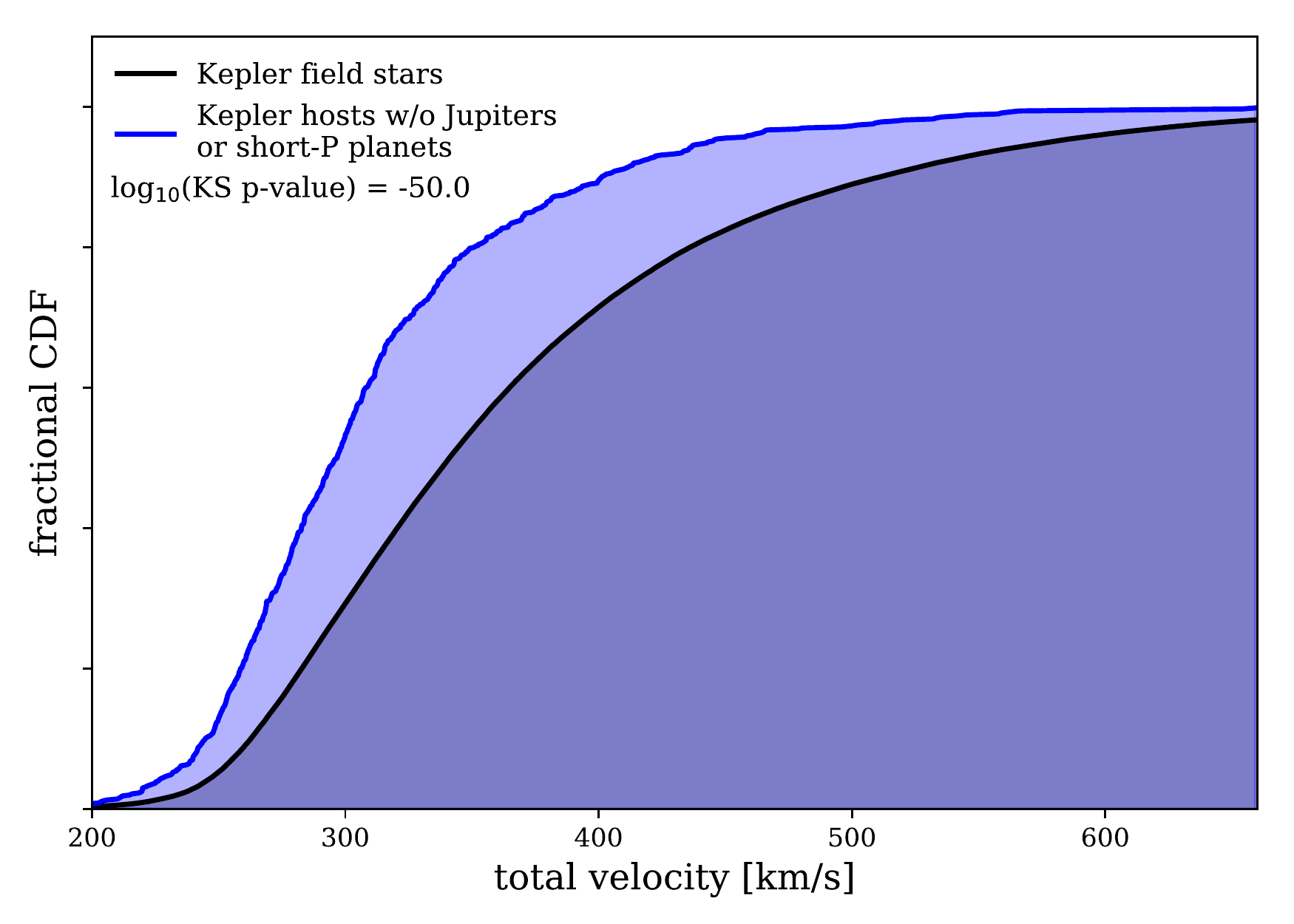}
\caption{CDF comparing the total velocities of \textit{Kepler} field stars (blue) and non-Jupiter hosts (black). Removing the Jupiter hosts doesn't have a strong effect on the velocity distribution of the sample, as it looks very similar to the distribution in Figure~\ref{fig:mot}.}
\label{fig:no_jup}
\end{figure}

The p-values were low enough in each velocity comparison that we can say with high confidence that \textit{Kepler} hosts and field stars are two distinct populations in terms of their motions. 

In the following sections, we explore potential explanations for this apparent distinction in velocity.

\subsection{Testing the planet-metallicity relation}
When we first saw the apparent difference in planet host and field star velocities, we considered that we may be indirectly picking up on the relation between stellar metallicity and planet occurrence rate. The relation posits that planets are more common around metal-rich stars \citep{wang:2015}, though the relation is strongest for Jupiter-mass planets and short-period planets (planets with periods shorter than 10 days) \citep{mulders:2016, petigura:2018, narang:2018}. We thought we might be picking up on this relation because metal-poor stars are more common in the thick disk of the Milky Way, where stellar motion is also higher.

To test whether or not our observation was a consequence of the planet-metallicity relation, we removed all of the Jupiter and short-period planet hosts from our host sample by eliminating stars which host planets larger than 0.5 Jupiter radii and planets with periods shorter than 10 days according to the \textit{Kepler} database. We call this sample the non-Jupiter sample for short. The remaining 700 hosts should have a weak correlation with stellar metallicity since they all host low-mass, intermediate-period planets.

Figure~\ref{fig:no_jup} shows the CDFs  comparing the total velocities of this non-Jupiter host sample and the field star sample. There is still a significant difference between the velocities of the two samples even with the hosts of Jupiters and short-period planets removed. From this test, we concluded that metallicity is not the main factor contributing to the difference between our samples' velocities.

\subsection{Testing thin vs. thick disk membership}\label{sec:disk}
Next we wondered if our observation was probing the dynamical difference between the thick and thin disks of the Milky Way. It's generally accepted that thick disk stars are more dynamically heated than thin disk stars, meaning they move more quickly. We wanted to be certain that our field star sample wasn't dominated by fast-moving thick disk stars. 

Though it's difficult to starkly distinguish the two disk populations, one common method is to separate them using vertical actions because the disks' dynamical heating is especially pronounced in the vertical direction \citep{soubiran:2003}.

\begin{figure}
\includegraphics[width=3in.]{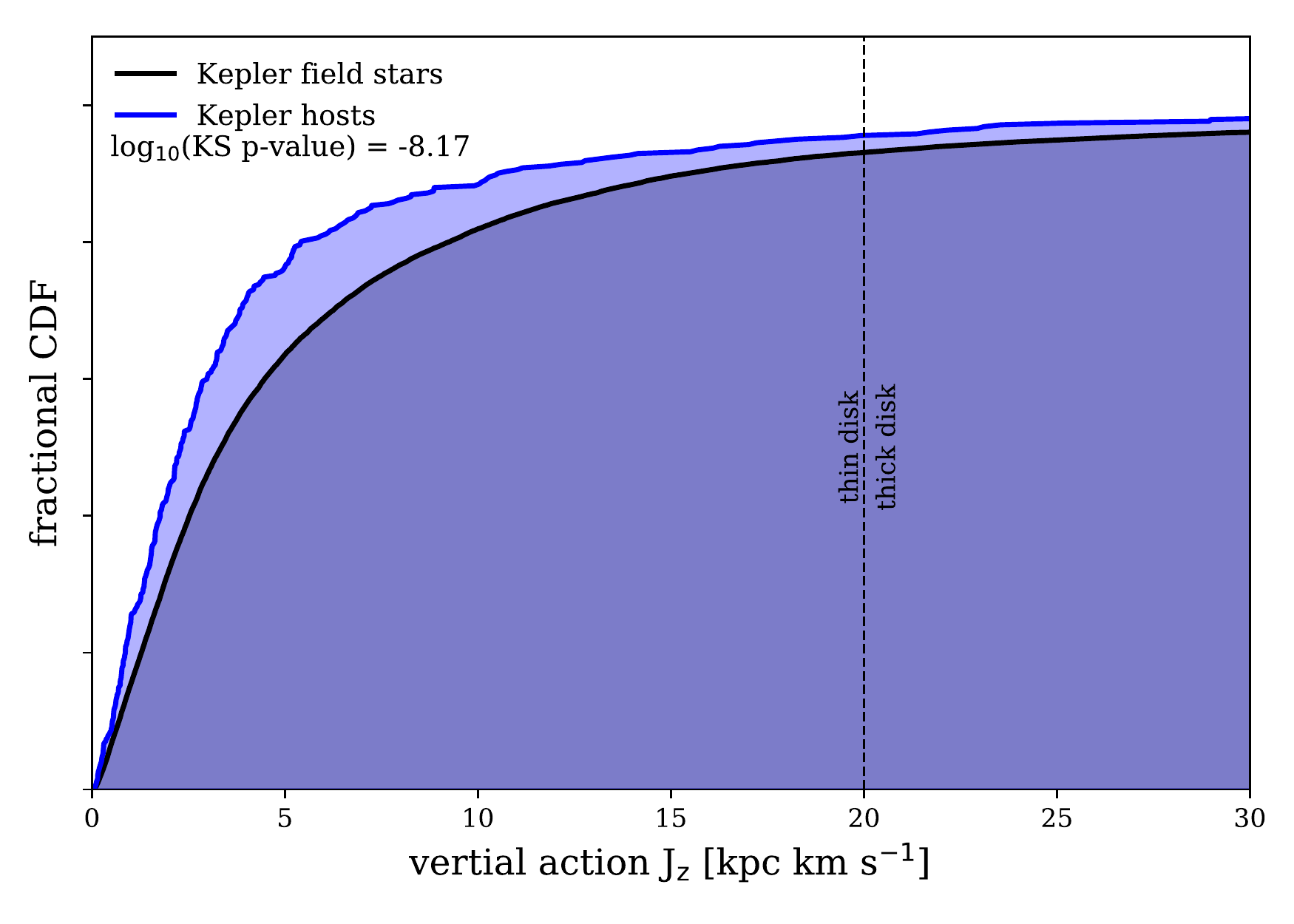}
\caption{Comparison of the vertical actions (J$_{\mathrm{z}}$) for \textit{Kepler} hosts (blue) and field stars (black) to see if the two populations can be said to belong to the thin disk and thick disk, respectively. Most stars in both samples have vertical actions consistent with thin disk membership.}
\label{fig:act}
\end{figure}

We used the vertical actions (J$_{\mathrm{Z}}$) calculated by \citet{sanders:2018} to compare the dynamics of our planet host sample to that of our field star sample. Figure~\ref{fig:act} shows the CDFs of our two samples' vertical actions.  

To understand which disk our samples come from, we adopted the J$_{\mathrm{Z}}$ cutoff of 20 kpc km/s from \citet{gandhi:2019}. Stars with J$_{\mathrm{Z}}$>20 kpc km/s are most likely part of the thick disk and stars with lower actions likely belong to the thin disk, though there's a $\sim$30\% contamination fraction. Almost all of the stars in both of our samples have J$_{\mathrm{Z}}$<10 kpc km/s, meaning our stars are likely all members of the Milky Way's thin disk and the difference in our samples' velocities is not a consequence of their disk membership.

\subsection{Testing \textit{Kepler} systematics}
Finally we checked to see if the difference between our samples' velocities was a consequence of \textit{Kepler}'s selection function. One possibility is that hotter stars, which are statistically younger \citep{fang:2017}, are moving slower by the velocity dispersion-age relation \citep{aumer:2016}. Hotter stars are brighter and thus more likely to have a planet detected by \textit{Kepler}. This would lead to an apparent but unreal difference in the velocities of field stars and (hot, young, slow) \textit{Kepler} hosts. 

\begin{figure}
\includegraphics[width=3.2in.]{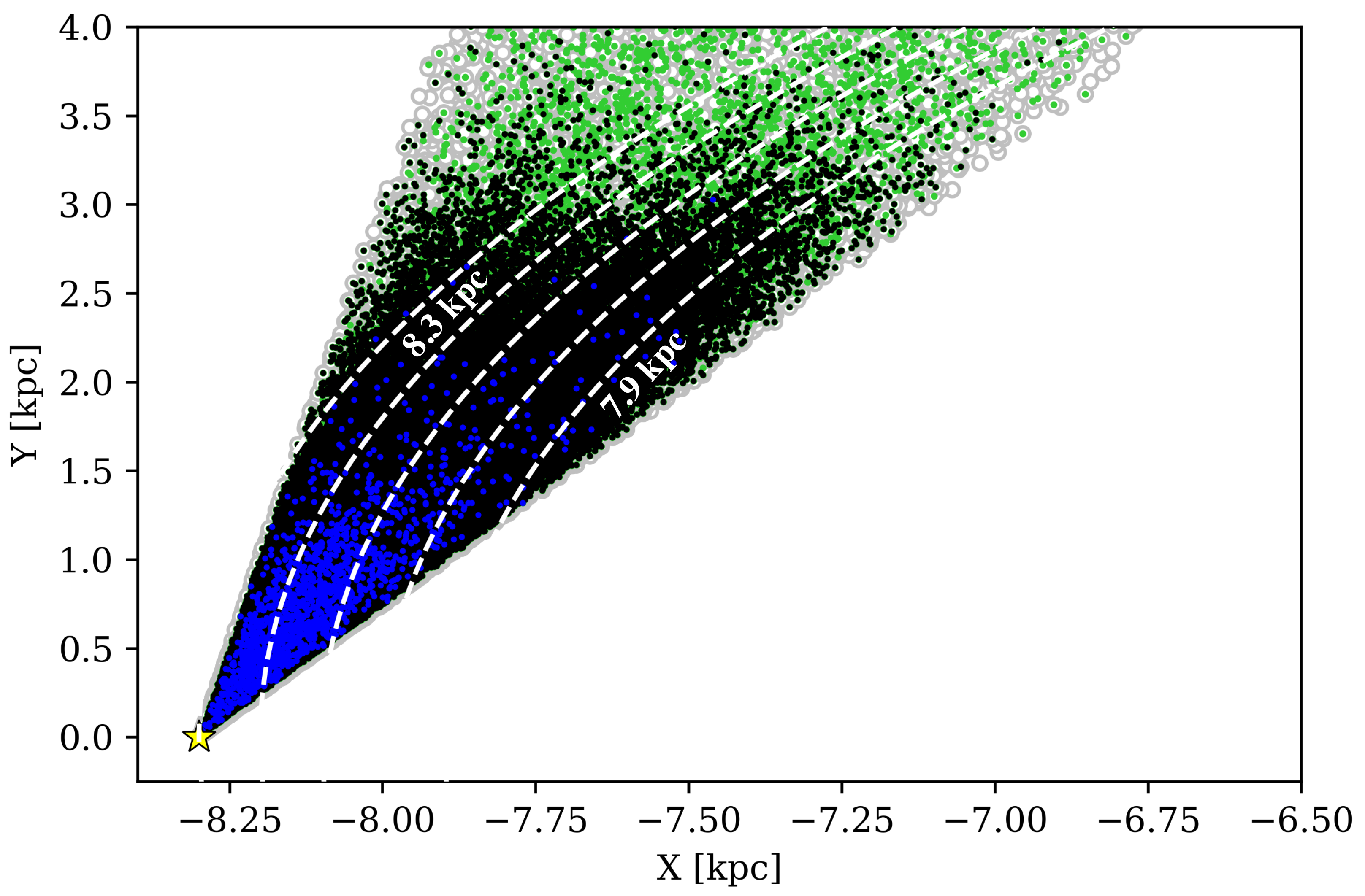}
\caption{Galactocentric X-Y positions for \textit{Kepler} hosts (blue) and field stars (black) with white rings indicating galactocentric radii}
\label{fig:pos_zoom}
\end{figure}

Another potential explanation is that \textit{Kepler} was more likely to find planets around closer stars, and the orientation of the \textit{Kepler} field of view makes it so that stars closer to the Sun are, on average, farther from the galactic center. (You can see this in Figure~\ref{fig:pos_zoom}, which shows a zoomed-in version of Figure~\ref{fig:data} with rings indicating galactocentric radii.) Therefore, confirmed \textit{Kepler} planet hosts could have slower galactocentric velocities than the ensemble of field star sample. We investigate these possibilities by finding stellar twins to all of the \textit{Kepler} hosts and examining their velocities.

Using distances from inverted \textit{Gaia} parallaxes along with stellar temperatures, surface gravities, and magnitudes from \textit{Kepler}, we created a sample made of stellar twins to the planet hosts. For each planet host, we found a star without any confirmed planets that had a distance, T$_{\mathrm{eff}}$, logg, and \textit{Kepler}-mag all within 5\% of the host's values. In doing this, we created a sample that's nearly identical to our planet host sample in every measurable way except for the fact that it doesn't include any confirmed planet hosts. Many of the planet hosts actually have thousands of stellar twins, so we produced 5000 twin samples and compared the velocities of each to those of our host and field stars samples.

If the host twins' distributions align with the field star sample, it would indicate that slow stellar velocity really is strongly linked to planet occurrence because the only similarity between twins and field stars is that they aren't confirmed planet hosts. But if the host twins' distributions align with the host sample, it would refute the planet-velocity relation because non-hosts are moving at the same speed as hosts. 

\begin{figure*}
\includegraphics[width=\textwidth]{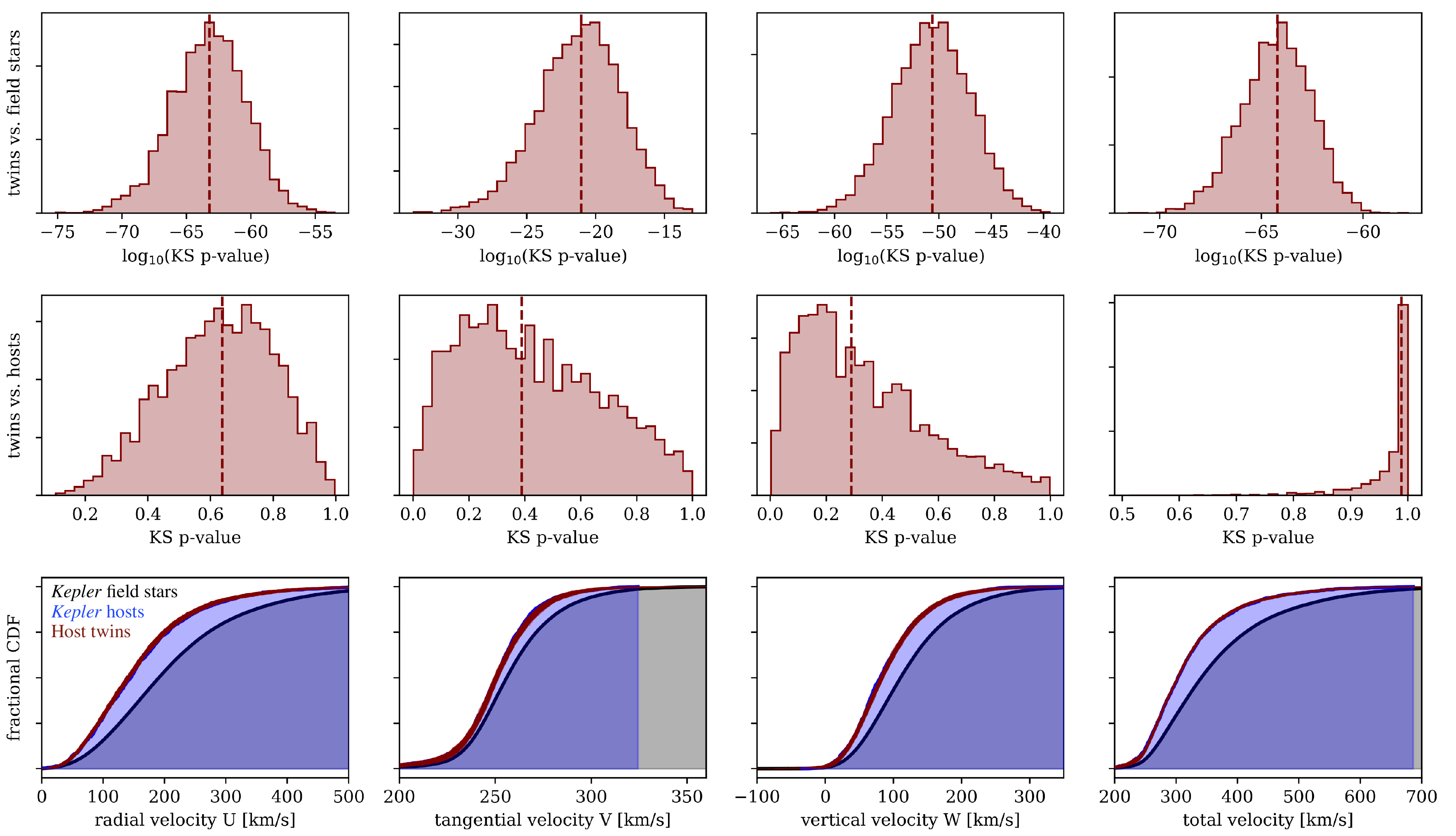}
\caption{\textbf{Top:} Distributions of KS p-values for our 5000 samples of host twins compared to the field star sample. \textbf{Middle:} Distributions of KS p-values for our 5000 samples of host twins compared to the planet hosts. \textbf{Bottom:} CDFs for the 5000 stellar twin samples are shown in red next to the velocity distributions for the \textit{Kepler} hosts (blue) and field stars (black). The twin distributions lie almost perfectly on top of the host distribution.}
\label{fig:twin_cdfs}
\end{figure*}

Figure~\ref{fig:twin_cdfs} shows the distributions of p-values calculated for all of our stellar twin samples compared to the \textit{Kepler} field stars as well as the p-values when compared to the planet hosts. The median p-values are indicated with a vertical dashed line. Figure~\ref{fig:twin_cdfs} also shows the CDFs of all 5000 stellar twin samples next to those of the \textit{Kepler} hosts and field stars to illustrate how those p-values in the top and middle rows translate to actual distributions. 

The CDFs of the stellar twins lie almost perfectly on top of the planet host distribution, even though the twins were selected without any velocity information.  

The similarity between the host sample and the ensemble of host twin samples suggests that the apparent difference between planet host and field star velocities can be completely explained by \textit{Kepler}'s selection function.

\section{Summary \& Discussion}\label{sec:summary}
In this work, we used \textit{Kepler} DR25 and \textit{Gaia} DR2 data to calculate the velocities of \textit{Kepler} hosts and field stars. We showed that \textit{Kepler} hosts have slower galactocentric velocities than the ensemble of \textit{Kepler} field stars, but we find that this difference can be completely explained as a consequence of \textit{Kepler}'s selection function. 

We were able to quantify the effect of this systematic bias as it relates to the planet-velocity relation by invoking a simple model. We assumed that the probability of a star with total galactocentric velocity $v_{\mathrm{tot}}$ hosting a planet can be described by an exponential distribution of form $v_0^{-1} e^{(-v_{\mathrm{tot}}/v_0)}$, where $v_0$ is a characteristic velocity above which the probability of hosting a planet drops to 1/$e$. This is motivated from a counting statistics framework. If the number of planets per star follows a homogeneous Poisson process, then the probability distribution of planet survival would be described by an exponential.

To infer the value of $v_0$, we used a Markov Chain Monte Carlo (MCMC) sampler with a uniform prior on $v_0$ and a likelihood set using a KS-test that compares a random twin sample's velocities to those of the \textit{Kepler} host sample. To 99\% confidence, $v_0 > $976 km/s, and to 99.9\% confidence, $v_0 > $623 km/s. Both of these values exceed the escape velocity of the Milky Way in the Sun's vicinity, which implies that planets in the Solar neighborhood are just as likely to form around high-velocity stars as they are around low-velocity stars.

When we first observed the apparent slowness of \textit{Kepler} hosts compared to field stars, we were reminded of the fact that few planets have been discovered in dense cluster environments \citep{fujii:2019}. If the dearth of planets in these dense environments is real and not another observational bias, it may be due to more frequent stellar encounters, which can destabilize planetary orbits. Clearly stellar encounters aren't common enough in the Solar neighborhood to have an effect on planet occurrence, but we are interested in exploring denser environments to see if the planet-velocity relation holds under such conditions. 

\section*{Acknowledgements}
This work made use of the gaia-kepler.fun crossmatch database created by Megan Bedell and the NASA Exoplanet Archive, which is operated by the California Institute of Technology, under contract with the National Aeronautics and Space Administration under the Exoplanet Exploration Program. M.A.S.M. is supported by the NSFGRF under grant No. DGE 16-44869. The authors wish to thank Melissa Ness, John Brewer, Devid Helfand, and members of the Cool Worlds Lab for helpful discussions.




\bibliographystyle{mnras}
\bibliography{mybibliography} 


\bsp	
\label{lastpage}
\end{document}